\documentclass[11pt]{article}
\usepackage{amssymb,amsmath,epsfig,graphics,verbatim,color,cite}
\begin{document}
\begin{sloppypar}
\vspace*{0cm}
\begin{center}
{\setlength{\baselineskip}{1.0cm}{ {\Large{\bf 
On integral and differential representations of Jordan chains and the confluent supersymmetry algorithm  \\}} }}
\vspace*{1.0cm}
{\large{\sc{Alonso Contreras-Astorga}$^\dagger$} and {\sc{Axel Schulze-Halberg}$^\ddagger$}}
\indent \vspace{.3cm} 
\noindent \\	
$~^{\dagger \ddagger}$Department of Mathematics and Actuarial Science, \\Indiana
University Northwest, 3400 Broadway, Gary IN 46408, USA,\\ 
$~^\ddagger$ Department of Physics, \\
Indiana
University Northwest, 3400 Broadway, Gary IN 46408, USA,\\
${}^\dagger$E-mail:
aloncont@iun.edu, alonso.contreras.astorga@gmail.com 
\\ ${}^\ddagger$E-mail:
axgeschu@iun.edu, xbataxel@gmail.com 
\end{center}

\vspace*{.5cm}
\begin{abstract}
\noindent

We construct a relationship between integral and differential representation of second-order Jordan chains. Conditions to obtain regular potentials through the confluent supersymmetry algorithm when working with the differential representation are obtained using this relationship. Furthermore,  it is used to find normalization constants of wave functions of quantum systems that feature energy-dependent potentials. Additionally, this relationship is used to express certain integrals involving functions that are solution of Schr\"odinger equations through derivatives.

\end{abstract} 

\section{Introduction}

A Jordan chain for a linear operator is a sequence of vectors that satisfies a particular system of equations. In the most elementary case, the linear operator is a square matrix and the corresponding Jordan chain is a sequence of generalized eigenvectors that are needed for the construction of Jordan's normal form  \cite{lopezgomez}. If the linear operator associated with the Jordan chain is a Hamiltonian, then the corresponding system of equations appears in the confluent version of the supersymmetry formalism (SUSY). The general purpose of SUSY is the construction of solvable quantum models (SUSY partners) and the manipulation of their spectra (spectral design). There is a large amount of literature on the topic, a selection of which can be found in the references of the reviews \cite{cooper,andrianov04} or the more recent work \cite{djf}. Within the formalism of SUSY, the standard algorithm and the confluent algorithm are to be distinguished. Being well understood, the vast majority of applications uses the standard algorithm, while its confluent counterpart is much less known. An introduction to the algorithm and discussion of mathematical properties can be found in \cite{stahlhofen95,baye95,boya98,mielnik00,djinv,xbatejp}  whereas some applications in  \cite{contreras14,contreras15,hussin15,grandati,correa15,bermudez15} and references therein. In the confluent SUSY algorithm, partner models are essentially determined through Jordan chains. The system of equations associated with a Jordan chain is solved by transformation functions that are used to construct the SUSY partner to a given model. The solution of such a system of equations can be represented in integral form \cite{xbatejp} or by means of derivatives \cite{bermudez}. While the process of obtaining the integral form is more straightforward \cite{djcon3,xbatejp}, a closed-form evaluation of the integrals is often not possible. Since the differential approach is much more feasible in calculations \cite{djcon1,djcon2}, it is of particular interest to find the mathematical relationship between the two approaches. In the present work, we determined the latter relationship and presented several applications regarding the confluent SUSY algorithm and the resolution of integrals to find normalization constants for energy dependent quantum mechanical systems.

The paper is organized as follows. We introduce and discuss 
Jordan chains of second order in section 2. We present in section 3 two main application of the results obtained: the confluent SUSY algorithm and energy dependent potentials, showing that some integrals appearing both situations can be expressed through derivatives. Finally, our conclusions are presented in the last section.

\section{Jordan chains of second order}
Let $D \subset \mathbb{R}$ be an open interval. The following system of differential equations is referred to as 
Jordan chain of second order
\begin{eqnarray}
u_{xx}(x,\lambda)+\left[\lambda-V(x,\lambda) \right]~u(x,\lambda) &=& 0, \label{j1} 
\\[1ex]
v_{xx}(x,\lambda)+\left[\lambda-V(x,\lambda) \right]~v(x,\lambda) &=& 
\left[V_\lambda(x,\lambda)-1 \right] u(x,\lambda). ~~~(x,\lambda) \in (D,\mathbb{R}).
\label{j2}
\end{eqnarray}
Here, the indices denote partial differentiation. We assume that the function $V$ and its partial 
derivative $V_\lambda$ are continuous. Furthermore, the solutions $u$ and $v$ are required to be 
three times continuously differentiable. Since equation (\ref{j1}) resembles the quantum-mechanical 
Schr\"odinger equation, the function $V$ and the variable $\lambda$ will be called potential and energy, 
respectively. In the particular case that the potential is independent of the energy, we have $V_\lambda=0$, such that 
the Jordan chain simplifies to
\begin{eqnarray}
u_{xx}(x,\lambda)+\left[\lambda-V(x) \right]~u(x,\lambda) &=& 0, \label{j1c} 
\\[1ex]
v_{xx}(x,\lambda)+\left[\lambda-V(x) \right]~v(x,\lambda) &=& -u(x,\lambda), ~~~(x,\lambda) \in (D,\mathbb{R}).
\label{j2c}
\end{eqnarray}
These equations play an important role in the 
confluent algorithm of the quantum-mechanical SUSY formalism, as will be discussed in detail below. 
Let us now return to our more general system (\ref{j1}), (\ref{j2}), where we are interested in the construction of 
solutions. Our starting point is the assumption that we know a solution $u$ of the first equation (\ref{j1}). From here 
we will distinguish two options for determining a solution $v$ to the second equation 
(\ref{j2}). 

\subsection{Integral representation} 

Our first option is the variation-of-constants formula, which requires two linearly independent 
solutions $u_1$ and $u_2$ of (\ref{j1}) in order to become applicable. For the sake of simplicity, let us 
make the following choices for these two functions
\begin{eqnarray}
u_1(x,\lambda) ~=~ u(x,\lambda) \qquad \qquad 
u_2(x,\lambda) ~=~ u(x,\lambda)~\int\limits^x \frac{1}{u^2(t,\lambda)}~dt. \label{u1u2}
\end{eqnarray}
Note that the integral form of $u_2$ is obtained through reduction of order \cite{kamke}. Observe further that due to the choice 
(\ref{u1u2}), the Wronskian $W_{u_1,u_2}$ of $u_1$ and $u_2$ 
is equal to one. The variation-of-constants formula then gives the following particular solution 
$v_{VC}$ to the second equation (\ref{j2}) of our Jordan chain.
\begin{eqnarray}
v_{VC}(x,\lambda) &=& \nonumber \\[1ex]
& & \hspace{-1.8cm} =~u_1(x,\lambda) \int\limits^x u(t,\lambda)~u_2(t,\lambda)~
\left[1-V_\lambda(t,\lambda) \right]dt - 
u_2(x,\lambda) \int\limits^x  u(t,\lambda)~u_1(t,\lambda)~\left[1-V_\lambda(t,\lambda) \right]dt, \nonumber
\end{eqnarray}
recall that the function $u$ appears in the nonhomogeneous term on the right side of (\ref{j2}). We can replace the 
latter function, as well as $u_2$, by their definitions shown in (\ref{u1u2}). We obtain 
the following expression
\begin{eqnarray}
v_{VC}(x,\lambda) &=& u(x,\lambda) \int\limits^x \left[\int\limits^x \frac{1}{u^2(t,\lambda)}~dt \right]u^2(t,\lambda) 
\left[1-V_\lambda(t,\lambda) \right]dt  \nonumber \\[1ex]
&-&
u(x,\lambda) \left[\int\limits^x \frac{1}{u^2(t,\lambda)}~dt \right] \left[\int\limits^x  u^2(t,\lambda) 
\left[1-V_\lambda(t,\lambda) \right]dt \right]. 
\nonumber 
\end{eqnarray}
After applying integration by parts to the first term on the right side, we arrive at the following integral representation of 
our solution $v_{VC}$
\begin{eqnarray}
v_{VC}(x,\lambda) &=& -  u(x,\lambda) \int\limits^x \left[\int\limits^t u^2(s,\lambda) \left[1-V_\lambda(s,\lambda) \right]ds 
\right] \frac{1}{u^2(t,\lambda)}~dt, \label{vvc}
\end{eqnarray}
The principal disadvantage of this representation is its limited applicability. Inspection of (\ref{vvc}) shows that 
the double integral can only be evaluated in closed form for relatively simple functions $u$ and $V$. For a solution $u$ of 
the second-order equation (\ref{j1}) that is typically given through special functions, it is in general not possible to 
evaluate (\ref{vvc}). 

\subsection{Differential representation} 

We will now derive a representation of the solution $v$ to (\ref{j2}) that does not 
contain any integral. To this end, we substitute the derivative $v=u_\lambda$ into the 
left side of (\ref{j1}). This gives 
\begin{eqnarray}
(u_{\lambda})_{xx}(x,\lambda)+\left[\lambda-V(x,\lambda) \right] u_\lambda(x,\lambda) &=& \nonumber \\[1ex]
& & \hspace{-5cm}
=~(u_{\lambda})_{xx}(x,\lambda) + \left\{\left[\lambda-V(x,\lambda) \right] u(x,\lambda) \right\}_\lambda
-\left[\lambda-V(x,\lambda) \right]_\lambda u(x,\lambda) \nonumber \\[1ex]
& & \hspace{-5cm}
=~\left\{
u_{xx}(x,\lambda)+\left[\lambda-V(x,\lambda) \right] u(x,\lambda)
\right\}_\lambda-\left[\lambda-V(x,\lambda) \right]_\lambda u(x,\lambda) \nonumber \\[1ex]
& & \hspace{-5cm}
=~-\left[\lambda-V(x,\lambda) \right]_\lambda u(x,\lambda). \label{j2check}
\end{eqnarray}
Note that in the last step we made use of (\ref{j1}). Next, we observe that (\ref{j2check}) is the same as our initial equation 
(\ref{j2}). Consequently, we have shown that 
\begin{eqnarray}
v_{DF}(x,\lambda)=u_\lambda(x,\lambda), \label{vdf} 
\end{eqnarray}
is a particular solution of (\ref{j2}), where the index is an abreviation for differential formula. 
Since (\ref{vdf}) is free of integrals, it is generally much easier to evaluate 
than (\ref{vvc}), and as such more suitable for applications.
\noindent \\ \\
While both the variation-of-constants scheme and the differential formula provide methods for solving (\ref{j2}), 
the respective particular solutions $v_{VC}$ and $v_{DF}$ are in general different from each other. It is therefore 
desirable to know the precise relation between these functions, such that one function can be expressed through its counterpart. 
In order to find this relation, let $u_1$ and $u_2$ be linearly independent solutions of (\ref{j1}), 
the Wronskian of which is equal to one. Recall that such a choice can be made through the settings (\ref{u1u2}). 
Since both $v_{VC}$ and $v_{DF}$ are particular solutions of the nonhomogeneous equation (\ref{j2}), their difference must be 
a solution to the homogeneous equation (\ref{j1}), which can be given as a linear combination of $u_1$ and $u_2$. Equivalently, 
there are two parameters $d_1$ and $d_2$, such that the equation
\begin{eqnarray}
d_1(\lambda)~u_{1}(x,\lambda)+d_2(\lambda)~u_{2}(x,\lambda) &=& v_{DF}(x,\lambda)-v_{VC}(x,\lambda), \label{eq1}
\end{eqnarray}
is fulfilled. The right side of this equation is completely determined by the choice of $u_1$ and $u_2$. Therefore, the remaining 
task is to calculate the functions $d_1$ and $d_2$. Since we need a second equation in 
order to find these two constants, we take the partial derivative of (\ref{eq1}) 
with respect to $x$, yielding
\begin{eqnarray}
d_1(\lambda)~(u_1)_x(x,\lambda)+d_2(\lambda)~(u_{2})_x(x,\lambda) &=& (v_{DF})_x(x,\lambda)-(v_{VC})_x(x,\lambda).  \label{eq2}
\end{eqnarray}
Hence, we now have an algebraic system of equations (\ref{eq1}), (\ref{eq2}) that is linear in $d_1$ and $d_2$. Application of 
Cramer's rule and taking into account that $W_{u_1,u_2}=1$, we arrive at the Wronskian representation of the solutions
\begin{eqnarray}
d_1(\lambda) ~=~ W_{v_{DF}-v_{VC},u_2}(x_0,\lambda),  \qquad \qquad \qquad
d_2(\lambda) ~=~ W_{u_1,v_{DF}-v_{VC}}(x_0,\lambda).   \label{d1d2}
\end{eqnarray}
Note that the constant $x_0 \in D$ can be chosen arbitrarily, since neither $d_1$ nor $d_2$ depend on the actual value of 
the variable $x$. In summary, if the functions $d_1$, $d_2$ are chosen as in (\ref{d1d2}), then the particular solutions 
 $v_{VC}$ and $v_{DF}$ of (\ref{j2}) are related to each other by means of the identity (\ref{eq1}). Recall that the explicit form of 
the particular solutions can be found in (\ref{vvc}) and (\ref{vdf}).

\section{Applications}
The Jordan chain and the differential representation of its solution has several interesting applications, two of which we will 
now introduce. The first application concerns the normalizability of solutions associated with Schr\"odinger equations for 
energy-dependent potentials. The second part of this section is devoted to the confluent SUSY formalism and 
several examples.

\subsection{The confluent SUSY algorithm}
The second application of our results on second-order Jordan chains concerns the 
confluent SUSY algorithm. For sake of completeness we present here a brief introduction to that algorithm, 
for a detailed description see \cite{djf} and references therein. 

We start out by considering the following stationary Schr\"odinger equation
\begin{eqnarray}
\psi_{xx}(x)+\left[\epsilon-V(x)\right] \psi(x)=0,
\end{eqnarray}
where $\epsilon$ is a real constant and the potential $V$ is a real function. Suppose further that we know two 
auxiliary functions $u,~v$ satisfying the Jordan chain
\begin{eqnarray}
u_{xx}(x,\lambda)+\left[\lambda-V(x) \right] u(x,\lambda) &=& 0, \label{Jordan1} \\
v_{xx}(x,\lambda)+\left[\lambda-V(x) \right] v(x,\lambda) &=& -u(x,\lambda), \label{Jordan2}
\end{eqnarray}
for a real constant $\lambda$. Observe that these equations are a special case of our general system (\ref{j1}), (\ref{j2}), where 
the potential does not depend on the variable $\lambda$. Then the function
\begin{eqnarray}
\phi(x,\lambda)~=~ \frac{W_{u,v,\psi}(x,\lambda)}{W_{u,v}(x,\lambda)}~=~ \frac{u^2(x,\lambda)}{W_{u,v}(x,\lambda)}~ \psi_x(x)+
\left[ \lambda - \epsilon - \frac{u(x,\lambda) ~u_x(x,\lambda)}{W_{u,v}(x,\lambda)} \right] \psi(x), \label{phi}
\end{eqnarray}
when $\epsilon \neq \lambda$, or in the opposite situation
\begin{eqnarray}
\phi(x,\lambda)=\frac{u(x,\lambda)}{W_{u,v}(x,\lambda)}, \label{missing}
\end{eqnarray}
fulfills the equation
\begin{eqnarray}
\phi_{xx}(x,\lambda)+\left[\epsilon-\widetilde{V}(x,\lambda)\right] \phi(x,\lambda)=0,
\end{eqnarray}
where the potential $\widetilde{V}$ is given by the expression
\begin{eqnarray}
\widetilde{V}(x,\lambda)= V(x) - 2 \left\{\log\left[W_{u,v}(x,\lambda) \right] \right\}_{xx}.  \label{newpot}
\end{eqnarray}
This transformation is known as confluent SUSY transformation and the potentials $\widetilde{V}$ and $V$ are called 
SUSY partners. Observe that the function $\phi$ in (\ref{phi}) depends on $\epsilon$. Since we are mainly interested in 
the dependence on $\lambda$, we do not include $\epsilon$ as an argument of $\phi$. 

Now, as can be seen from \eqref{phi}, \eqref{missing} and \eqref{newpot}  the Wronskian $W_{u,v}$ plays a 
fundamental role to perform the transformation. Furthermore, to obtain regular potentials $\widetilde{V}$ 
we should avoid zeros in this Wronskian. If the transformation function $v$ is obtained by the variation-of-constants 
method (see \eqref{vvc}), then $W_{u,v}$ can be expressed as 
\begin{eqnarray}
W_{u,v_{VC}}(x,\lambda)=\omega_0 - \int^x_{x_0} u^2(t,\lambda) dt,  \label{W int}
\end{eqnarray}
where $x_0$ is a point in the domain of $V$ and $\omega_0$ is an arbitrary constant. The conditions for 
obtaining a regular potential $\widetilde{V}$ from the confluent SUSY transformation are known and can be 
presented as follows: Let the domain of $V$ be the interval $(x_\ell, x_r)$, then we have 
two non-excluding options:
\begin{itemize}
\item We use a solution of \eqref{Jordan1} as transformation function such that $u(x_\ell)=0$. In this case the integral 
\begin{eqnarray}
I_\ell \equiv  \int^{x_0}_{x_\ell} u^2(y,\lambda) dy < \infty, 
\end{eqnarray} 
and as a consequence if $\omega_0 \in (-\infty, -I_\ell]$, we avoid zeros in the Wronskian.

\item The transformation function $u$ satisfies $u(x_r)=0$. Thus, the integral 
\begin{eqnarray}
I_r \equiv \int^{x_r}_{x_0} u^2(y,\lambda) dy < \infty, 
\end{eqnarray}
and if $\omega_0 \in [I_r,\infty )$, we guarantee that $W_{u,v_{VC}}$ never vanishes.
\end{itemize}
  
When the solution $v$ of  \eqref{Jordan2} is a superposition of the general solution of the corresponding homogeneous equation and the particular solution obtained by differentiating $u$ with respect to the parameter 
$\lambda$ (see \eqref{vdf}) the regularity conditions have not been fully explored. Let us express 
the Wronskian by
\begin{eqnarray}
W_{u,v}(x,\lambda)&=&K+W_{u,u_{\lambda}}(x,\lambda), \label{W diff}
\end{eqnarray}
where $K$ is an arbitrary constant. A derivative with respect to the position, and using the Jordan chain \eqref{Jordan1}--\eqref{Jordan2} to simplify second order derivatives, leads to 
\begin{eqnarray}
\frac{\partial W_{u,v}(x,\lambda)}{\partial x}= - u^2(x,\lambda), \label{dWdx}
\end{eqnarray}
i.e.,  for a fixed $\lambda$, $W_{u,v}$ is a non increasing monotone function of $x$. Thus,  two conditions can lead to a regular potential: 

\begin{itemize}
\item Non diverging transformation function $u(x,\lambda)$ when $x \rightarrow x_\ell$ and 
\begin{eqnarray}
K \leq - W_{u,u_{\lambda}}(x_\ell,\lambda). \label{conditon dif 1}
\end{eqnarray}

\item Non diverging transformation function $u(x,\lambda)$ when $x \rightarrow x_r$ and 
\begin{eqnarray}
K \geq - W_{u,u_{\lambda}}(x_r,\lambda). \label{condition dif 2}
\end{eqnarray}

\end{itemize}

Note that when the transformation function fulfills both boundary conditions, 
the domain of $K$ is $(-\infty, - W_{u,u_{\lambda}}(x_\ell,\lambda)]  \cup [ - W_{u,u_{\lambda}}(x_r,\lambda), 
\infty )$.

Furthermore, integration of \eqref{dWdx} leads to the interesting result
\begin{eqnarray}
\int_{x_0}^x u^2(t,\lambda) dt = W_{u,u_{\lambda}}(x_0,\lambda) - W_{u,u_{\lambda}}(x,\lambda),  \label{integral}
\end{eqnarray}
where $x_0 \in (x_\ell,x_r)$. This last equation shows how to integrate the square of a function that is solution of a 
Schr\"odinger equation using a Wronskian and vice versa. This identity is useful when we are interested, for example, 
in finding normalization constants or probabilities in an interval.

In the following two examples we illustrate first with the infinite well quantum system how to obtain the normalization 
constant of its eigenfunctions using \eqref{integral} and then how to obtain its SUSY partners using 
$v_{DF}$ and the regularity conditions presented in \eqref{conditon dif 1} and \eqref{condition dif 2}. Since 
SUSY partners of this potential have been obtained in \cite{fernandez07} using the representation of the 
Wronskian as in \eqref{W int}, this example provides a good comparison of the two methods. In the second example 
we build new exactly solvable potentials departing from the radial oscillator system and study the regularity conditions 
for obtaining these potentials.

\subsubsection{Particle in a box}
Consider a particle included between two impenetrable potential walls at $x=0$ and $x=1$. The Schr\"odinger equation of the system is 
\begin{eqnarray}
\psi_{xx}(x)+\epsilon ~\psi(x) = 0, \qquad x \in (0,1),  \label{box} 
\end{eqnarray}
and the boundary conditions are
\begin{eqnarray}
\lim_{x \rightarrow 0} \psi(x,\epsilon) = \lim_{x \rightarrow 1} \psi(x,\epsilon) = 0. \label{box boundary}
\end{eqnarray}
If $\epsilon=k^2$ the the general solution of \eqref{box} is given by
\begin{eqnarray}
\psi(x) = A \sin\left(k x\right)+B \cos\left(k x\right) \label{box general}
\end{eqnarray}
where $A$ and $B$ are arbitrary constants. To satisfy the boundary conditions \eqref{box boundary} we need to set $B=0$ and 
$k = n \pi$, where $n$ is a natural number, then the eigenfunctions and eigenvalues are 
\begin{eqnarray}
\psi(x) = A \sin\left(n \pi x \right),  \qquad \epsilon = n^2 \pi^2,  \label{box eigen}
\end{eqnarray}
the normalization constant $A$ of which we can obtain by means of \eqref{integral}. In order to avoid confusion regarding 
the notation, we extend the solution (\ref{box general}) to a function of two variables $\psi=\psi(x,\epsilon)$, where it is 
understood that $|k|=\sqrt{\epsilon}$. The derivative with respect to $\epsilon$ can be obtained using the chain rule
\begin{eqnarray}
\psi_\epsilon(x,\epsilon) = \frac{\partial k}{\partial \epsilon } \frac{\partial \psi}{\partial k} = \frac{A x}{2 k} \sin\left(k x \right), \label{parder}
\end{eqnarray}
note that we did not include the variable of differentiation $\epsilon$ as an argument of $\psi$. We have 
\begin{eqnarray}
W_{\psi, \psi_\epsilon}(x,\epsilon)= \frac{A^2}{2} \left[ \frac{\sin\left(2 k x\right)}{2 k} - x^2 \right],
\end{eqnarray}
finally, the normalization condition leads us to 
\begin{eqnarray}
1=\int_0^1 \psi^2 dx = W_{\psi, \psi_\epsilon}(0,\epsilon)-W_{\psi, \psi_\epsilon}(1,\epsilon)= \frac{A^2}{2}  \quad \Rightarrow \quad  A= \sqrt{2}.
\end{eqnarray}

\begin{figure}[t]
\begin{center}
\includegraphics[width=7.5 cm]{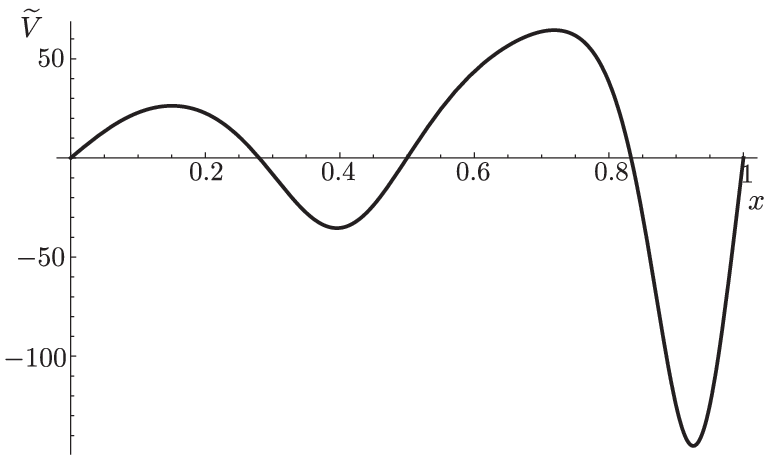}  \hspace{.4cm}
\includegraphics[width=7.5 cm]{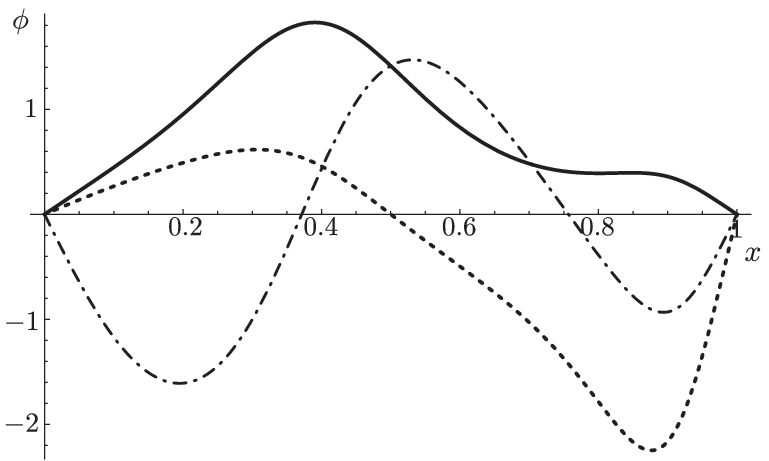}
\caption{On the left, potential $\widetilde{V}$ using $\lambda=4 \pi^2$ and $K=0.555$. On the right its first three eigenfunctions.}
\label{Fig Box}
\end{center}
\end{figure}

To generate a exactly solvable potential through the confluent SUSY formalism using the differential representation 
of the Wronskian \eqref{W diff} set $\lambda= m^2 \pi^2$, where $m$ is a natural number, then the transformation functions $u$ and $v$ satisfying the Jordan chain \eqref{Jordan1}-\eqref{Jordan2} for $V(x)=0$. Let us consider 
\begin{eqnarray}
u(x,\lambda) &=&  \sin\left(\sqrt{\lambda} x \right), \label{u box} \\
v(x,\lambda) &=& \frac{1}{2 \sqrt{\lambda}} ~x \cos\left(\sqrt{\lambda} x \right) - \frac{K}{\sqrt{\lambda}}\cos\left(\sqrt{\lambda} x\right), \label{v box}
\end{eqnarray}
where $K$ is an arbitrary constant. Note that the first term of $v$ is a particular solution of \eqref{Jordan2} obtained as 
$v_{DF}=u_\lambda$, while the second term is solution of the corresponding homogeneous equation. Now, to obtain regular 
potentials we need to evaluate the Wronskian $W_{u,u_\lambda}$ in $x=x_\ell$ and $x=x_r$:  
\begin{eqnarray}
W_{u,u_\lambda}(0,\lambda)= 0, \qquad W_{u,u_\lambda}(1,\lambda) = -\frac{1}{2}. 
\end{eqnarray}
According to the conditions \eqref{conditon dif 1} and \eqref{condition dif 2}, if the constant $K \in (-\infty,0] \cup [1/2, \infty)$ then the 
produced potential is regular. The potential generated is given by \eqref{W diff} and can be expressed as 
\begin{eqnarray}
\widetilde{V}(x,\lambda)= \frac{16 \pi^2 m^2 \left[1+m \pi(2 K-x)\sin(2m \pi x)-\cos(2 m \pi x)  \right]}{\left[2 m \pi(2 K - x)+\sin (2 \pi  m x) \right]^2}, \label{boxSUSYP}
\end{eqnarray}
recall that $m= \sqrt{\lambda}/\pi$. Figure \ref{Fig Box} shows a potential $\widetilde{V}$ (see \eqref{boxSUSYP}) where transformation functions \eqref{u box}--\eqref{v box} with a parameter $\lambda=4 \pi^2$ were used. On the right, it can be seen its first three eigenfunctions (continuous, dotted and dashed curves respectively), the first and third were generated with the rule \eqref{phi} while the second with \eqref{missing} since $\lambda=4 \pi^2$  is also and eigenvalue of the original system, see \eqref{box eigen}.    

Moreover, any integral of the form $\int u^2 dx$, where $u$ is solution of \eqref{Jordan1}, can be obtained with 
\eqref{integral} even though the boundary condition \eqref{box boundary} are not satisfied. Let us consider 
\begin{eqnarray}
u_1(x,\lambda)= \sin \left(\sqrt{\lambda} x \right), \qquad u_2 (x,\lambda)= - \frac{1}{\sqrt{\lambda}} \cos\left( \sqrt{\lambda} x \right), \label{u1u2 box}
\end{eqnarray}
then the following integrals can be obtained with \eqref{integral}:
\begin{eqnarray}
\int_0^x u_1^2(t,\lambda)dt&=& W_{u_1, (u_1)_\lambda}(0,\lambda)-W_{u_1, (u_1)_\lambda}(x,\lambda)=\frac{x}{2}-\frac{1}{4\sqrt{\lambda} }\sin \left(2 \sqrt{\lambda} x \right), \label{int1 box} \\
\int_0^x u_2^2(t,\lambda)dt&=& W_{u_2, (u_2)_\lambda}(0,\lambda)-W_{u_2, (u_2)_\lambda}(x,\lambda)=\frac{x}{2\lambda}+\frac{1}{4 \lambda^{3/2} }\sin \left(2 \sqrt{\lambda} x \right). \label{int2 box}
\end{eqnarray}
Furthermore, using the relations  \eqref{eq1} and \eqref{d1d2} doubles integrals of the form $\int (\int u^2 dx)/u^2 dx$ can be found. The transformation function $v$ using the variation of constant formula \eqref{vvc} in this example is given by
\begin{eqnarray}
v_{VC}(x,\lambda)= - u_1(x,\lambda) \int_{x_0}^x\left[\int_{x_0}^t u_1^2(s,\lambda) ds\right] \frac{1}{u_1^2 (t,\lambda)}dt,
\end{eqnarray}  
notice that the two integration constants associated with the integrals have been fixed by means of the integration limits, where $x_0$ is an arbitrary number in the domain $D$ of our Jordan chain. Now, the function $v_{VC}$ satisfies the conditions
\begin{eqnarray}
v_{VC}(x_0,\lambda)=(v_{VC})_x(x_0,\lambda)=0.
\end{eqnarray}
Keeping this in mind, the formulas in \eqref{d1d2} reduce to 
\begin{eqnarray}
d_1(\lambda)= W_{v_{DF}, u_2}(x_0,\lambda), \qquad d_2(\lambda)= W_{u_1,v_{DF}}(x_0,\lambda).  \label{a d1d2}
\end{eqnarray}
Let us point out that the functions in \eqref{a d1d2} do not depend on any integrals. Now, solving \eqref{eq1} for the double integral in it,  
\begin{eqnarray}
\int_{x_0}^x\left[\int_{x_0}^t u_1^2(s,\lambda) ds\right] \frac{1}{u_1^2 (t,\lambda)}dt = - \frac{1}{u_1(x,\lambda)} \left[(u_1)_\lambda (x,\lambda)-d_1(\lambda)u_1-d_2(\lambda) u_2(x,\lambda)   \right],   \label{double int}
\end{eqnarray}
recall that $v_{DF}=(u_1)_\lambda$. Substitution of \eqref{u1u2 box} in \eqref{double int} leads to 
\begin{eqnarray}
\int_{x_0}^x\left[\int_{x_0}^t u_1^2(s,\lambda) ds\right] \frac{1}{u_1^2 (t,\lambda)}dt =   \frac{\cos^2\left(\sqrt{\lambda} x_0 \right)}{2 \lambda}- \left[ \frac{x-x_0}{2 \sqrt{\lambda}} +\frac{\sin \left(2 \sqrt{\lambda} x_0 \right)}{4 \lambda}  \right] \cot \left(\sqrt{\lambda} x \right).  \label{double int a}
\end{eqnarray}
In this simple case, integrals in \eqref{int1 box}, \eqref{int2 box} and \eqref{double int a} can be verified by direct integration.

\subsubsection{The radial oscillator system}

\begin{figure}[t]
\begin{center}
\includegraphics[width=7.5 cm]{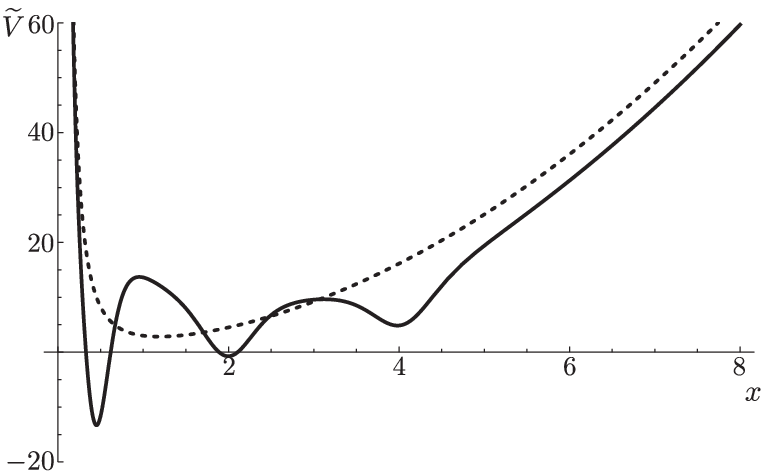}  \hspace{.4cm}
\includegraphics[width=7.5 cm]{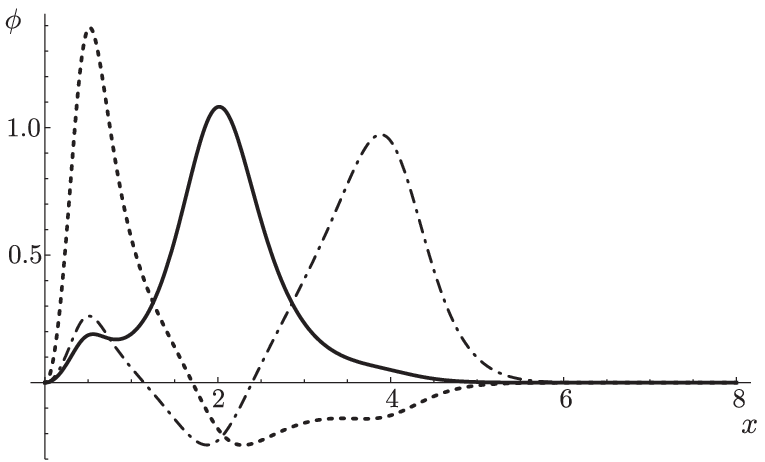}
\caption{On the left, a potential $\widetilde{V}$ using a transformation function $u$ with the parameters $\ell=1,~ \lambda=8$ and $K=-0.01$ (continuous curve) and as a reference the potential $V$ (dotted curve). On the right its first three eigenfunctions.}
\label{Fig Osc}
\end{center}
\end{figure}

The Sch\"odinger equation for the radial oscillator potential is given by \cite{flugge}
\begin{eqnarray}
\psi_{xx}(x) + \left[\epsilon -x^2 - \frac{\ell(\ell+1)}{x^2}  \right] \psi(x) = 0, \qquad x\in (0,\infty), \label{oscillator}
\end{eqnarray}
where the potential $V$ can be identified as 
\begin{eqnarray}
V(x)= x^2 +  \frac{\ell(\ell+1)}{x^2}, \label{osc pot}
\end{eqnarray}
and we will consider in this example $\ell$ as a natural number. The boundary conditions of this quantum problem are
\begin{eqnarray}
\lim_{x \rightarrow 0} \psi(x) = \lim_{x \rightarrow \infty} \psi(x) = 0. \label{radial boundary}
\end{eqnarray}
The general solution of \eqref{oscillator} without considering the boundary condition is 
\begin{eqnarray}
\psi (x) &=& x^{\ell+1} \exp\left(-\frac{x^2}{2} \right) \left\{ A~_1F_1 \left( \frac{1}{4} \left(2 \ell +3-\epsilon \right); \ell+\frac{3}{2}; x^2 \right)  \right. \nonumber \\
& & + \left. B~ x^{-(2 \ell +1)} ~_1F_1 \left( \frac{1}{4}\left(-2\ell+1-\epsilon \right); -\ell+\frac{1}{2}; x^2\right)    \right\}, \label{osc general}
\end{eqnarray}
where $_1F_1$ is the confluent hypergeometric function \cite{abram}, and $A$, $B$ are arbitrary constants. Since the second term of $\psi$ as given in \eqref{osc general} is divergent at the origin we need to set $B=0$ in order to satisfy the first boundary condition. The second condition is fulfilled only when the first argument of the hypergeometric function is a negative integer. Thus, the eigenfunctions and eigenvalues of \eqref{oscillator} are 
\begin{eqnarray}
\psi(x) = A~ x^{\ell+1} \exp\left(-\frac{x^2}{2} \right) ~_1F_1 \left( -n ; \ell+\frac{3}{2}; x^2 \right), \quad \epsilon = 4n +2\ell +3, 
\quad  n= 1, 2, 3, \dots, \label{eigenvalue osc}
\end{eqnarray}
To obtain the confluent SUSY partners \eqref{newpot} of the potential \eqref{osc pot} using the differential formula \eqref{W diff} 
we need a solution of the Jordan chain \eqref{Jordan1}--\eqref{Jordan2}. We will use the solution $u$ of the first equation in the Jordan chain as given in \eqref{osc general} with $A=1$, $B=0$ and $\lambda$ different from any of the  eigenvalues: 
\begin{eqnarray}
u(x,\lambda)= x^{\ell+1} ~\exp \left(- \frac{x^2}{2} \right) ~_1F_1 \left( \frac{1}{4}(2 \ell +3 -\lambda); \ell + \frac{3}{2}; x^2  \right). \label{osc u}
\end{eqnarray}
A particular solution of the second equation in the Jordan chain can be constructed as $u_\lambda$. 
In order to obtain this derivative, the formula of the derivatives with respect the parameters of the confluent hypergeometric 
function can be found in \cite{ancarani08}. Then the second transformation function can be expressed as:
\begin{eqnarray}
v(x,\lambda)&=& -\frac{1}{4} x^{\ell+1} ~\exp \left(- \frac{x^2}{2} \right) \left[ \sum_{m=0}^\infty \frac{(\frac{1}{4}(2 \ell +3 -\lambda))_m}{(\ell + \frac{3}{2})_m} \frac{x^{2m}}{m !} \sum_{p=0}^{m-1} \frac{1}{p+\frac{1}{4}(2 \ell +3 -\lambda)} \right] \nonumber  \\
& &-  \frac{K}{2\ell+1}~ x^{-\ell} \exp\left(-\frac{x^2}{2} \right) ~_1F_1 \left( \frac{1}{4}\left(-2\ell+1-\lambda \right); -\ell+\frac{1}{2}; x^2\right), \label{osc v}    
\end{eqnarray}
where $(\cdot)_m$ is the Pochhammer symbol \cite{abram} and $K$ is an arbitrary constant. Note that the first term in \eqref{osc v} is the particular solution while the second is solution of the homogeneous equation.  
To simplify some expressions let us define an auxiliary function $h=h(x,\lambda)$ as:
\begin{eqnarray}
h(x,\lambda)= \frac{(2\ell +3-\lambda)}{(2 \ell +3)} x^{\ell+2} ~\exp \left(- \frac{x^2}{2} \right) ~_1F_1 \left( \frac{1}{4}(2 \ell +7 -\lambda); \ell + \frac{5}{2}; x^2  \right).
\end{eqnarray}
Notice that the expression of the Wronskian $W_{u,u_\lambda} = uh_\lambda - u_\lambda h$. 
The expression of the SUSY partners $\widetilde{V}$ is in \eqref{newpot} and for the radial oscillator can be expressed as  
\begin{eqnarray}
\widetilde{V}(x)= x^2+\frac{\ell(\ell+1)}{x^2}+ \frac{4 u(x,\lambda) u_x(x,\lambda) \left[ K+ u(x,\lambda) h_\lambda(x,\lambda)  - u_\lambda(x,\lambda) h(x,\lambda) \right]+2 u^4(x,\lambda)}{\left[ K+ u(x,\lambda) h_\lambda(x,\lambda)  - u_\lambda(x,\lambda) h(x,\lambda) \right]^2 }.  \label{osc susypot}
\end{eqnarray} 
According to the condition \eqref{conditon dif 1}, and since  $W_{u,u_\lambda}(0,\lambda)=0$, to obtain regular potentials the value of the constant $K$ has to be a non positive real number. On the left of figure \ref{Fig Osc} it can be seen a potential $\widetilde{V}$, see \eqref{osc susypot}, and as a reference the radial  oscillator potential \eqref{osc pot} (dotted curve). The parameters used are  $\ell=1,~ \lambda=8$ and $K=-0.01$. On the right we plotted its first three eigenfunctions, the first and third (continuous and dashed curves, respectively) were obtained with the rule \eqref{phi} while the second (dotted curve) with \eqref{missing}. 

Before we conclude this section, let us comment on one more application that arises within the present example of the 
radial oscillator system. As we observe in (\ref{osc general}), the general solution of equation (\ref{oscillator}) is a linear combination of 
two solutions that we will call $u_1$ and $u_2$. Using the abbreviations
\begin{eqnarray}
a=\frac{1}{4}\left(2 \ell + 3 -\lambda \right),\quad b= \ell+\frac{3}{2}, \quad c=\frac{1}{4} \left(-2\ell +1 -\lambda  \right), \quad 
d=-\ell + \frac{1}{2},
\end{eqnarray}
the particular solutions in (\ref{osc general}) can be written as
\begin{eqnarray}
u_1(x,\lambda)&=& x^{\ell+1} \exp \left(-\frac{x^2}{2}\right)~_1F_1\left(a;b;x^2 \right),  \\
u_2(x,\lambda)&=&- \frac{1}{2\ell+1}~x^{-\ell} \exp \left(-\frac{x^2}{2}\right)~_1F_1\left(c;d;x^2 \right). 
\end{eqnarray}
Since $u_1$ and $u_2$ are solutions of equation \eqref{oscillator} for $\epsilon = \lambda$,  the identity \eqref{integral} can be used 
to find several integrals involving the latter functions. We obtain
\begin{eqnarray}
\int_0^x u_1^2(t,\lambda) dt &=&   \frac{a  }{2 b} x^{2 \ell +3} \exp \left(-x^2\right)  \left\{\, _1F_1\left(a;b;x^2\right) 
\sum _{n=0}^{\infty } \frac{ (a+1)_n~ x^{2 n}}{(b+1)_n ~ n! }  \sum _{p=0}^{n-1} \frac{1}{p+a+1} \right.  \nonumber  \\
&+&\left.  _1F_1\left(a+1;b+1;x^2\right) \left[ \frac{\, _1F_1\left(a;b;x^2\right)}{a}-\sum _{n=0}^{\infty } \frac{(a)_n~x^{2 n}}{ (b)_n ~n!} 
\sum _{p=0}^{n-1} \frac{1}{p+a}\right] \right\},
\end{eqnarray}
and 
\begin{eqnarray}
\int_x^\infty u_2^2(t,\lambda) dt &=& -\frac{c ~ x^{1-2 \ell } }{2(2\ell+1)^2 d}\exp \left(-x^2\right)  \left\{\, _1F_1\left(c;d;x^2\right) 
\sum _{n=0}^{\infty } \frac{ (c+1)_n ~x^{2 n}}{(d+1)_n~n!} \sum _{p=0}^{n-1} \frac{1}{p+c+1} \right. \nonumber \\
& & + \left. _1F_1\left(c+1;d+1;x^2\right) \left[\frac{\, _1F_1\left(c;d;x^2\right)}{c}-\sum _{n=0}^{\infty } \frac{(c)_n ~x^{2 n}}{ (d)_n ~n!} 
\sum _{p=0}^{n-1} \frac{1}{p+c} \right]\right\}. 
\end{eqnarray}
Furthermore, since $u=u_1+u_2$ is also solution of the same differential equation and due to the linearity of the Wronskian we can 
 obtain the integral of the the product $u_1 u_2$. This results in the comparatively long expression
\begin{eqnarray}
\int_{x_0}^x u_1(t,\lambda)u_2(t,\lambda) dt &=& \frac{1}{8(2\ell+1)} \left\{(2 \ell +1) \exp \left(-x^2\right) \left[\, _1F_1\left(a;b;x^2\right) 
\sum _{n=0}^{\infty } \frac{(c)_n ~x^{2 n}}{(d)_n ~n!}\sum _{p=0}^{n-1} \frac{1}{p+c} \right. \right. \nonumber \\
&-& \left. \left.  _1F_1\left(c;d;x^2\right) \sum _{n=0}^{\infty } \frac{(a)_n~x^{2 n}}{(b)_n ~n!}\sum _{p=0}^{n-1}
 \frac{1}{p+a}\right] \right. \nonumber \\
&+& \left. 2 x^2~ \exp \left(-x^2\right) \left[-\frac{c}{d} \, _1F_1\left(a;b;x^2\right) \sum _{n=0}^{\infty } \frac{(c+1)_n ~
x^{2 n}}{(d+1)_n~n!} \sum _{p=0}^{n-1} \frac{1}{p+c+1} \right. \right. \nonumber \\ 
&-& \left. \left. \frac{a}{b} \, _1F_1\left(c;d;x^2\right) \sum _{n=0}^{\infty } \frac{(a+1)_n~x^{2 n}}{(b+1)_n~n!}
\sum _{p=0}^{n-1} \frac{1}{p+a+1} \right. \right. \nonumber \\
&+& \left. \left. \frac{c}{d} \, _1F_1\left(c+1;d+1;x^2\right) \sum _{n=0}^{\infty } \frac{(a)_n~x^{2 n}}{(b)_n~n!}
\sum _{p=0}^{n-1} \frac{1}{p+a} \right. \right. \nonumber \\
&+&\left. \left.  \frac{a}{b} \, _1F_1\left(a+1;b+1;x^2\right) \sum _{n=0}^{\infty } \frac{(c)_n~ x^{2 n} }{ (d)_n~n!}
\sum _{p=0}^{n-1} \frac{1}{p+c} \right. \right. \nonumber \\ 
&-&\left. \left. \left. \frac{1}{d} \, _1F_1\left(a;b;x^2\right) \, _1F_1\left(c+1;d+1;x^2\right) \right. \right. \right. \nonumber \\
&-& \left. \left. \left.  \frac{1}{b} \, _1F_1\left(a+1;b+1;x^2\right) \, _1F_1\left(c;d;x^2\right)\right]\right\} \right|_{x_0}^x.
\end{eqnarray}
While this expression is very difficult to manage by hand, it can easily be processed further through a symbolic calculator.

\subsection{Energy-dependent potentials and normalization}
It is well-known that quantum models involving energy-dependent interactions are subject to a modified 
theory \cite{formanek}. In particular, the usual completeness relation and the corresponding $L^2$-norm 
of wavefunctions associated with energy-dependent potentials change their form. If we interpret 
(\ref{j1}) as a Schr\"odinger equation for an energy-dependent potential $V$ that admits a bound state solution 
$u$, then the norm $N(u)$ of such a solution must be calculated as follows \cite{formanek}
\begin{eqnarray}
N(u) &=& \int\limits_D \left[1-V_\lambda(t,\lambda) \right] | u(t,\lambda) |^2~ dt. \label{enorm}
\end{eqnarray}
Here we use the symbol $N$ rather than the usual notation $\Vert \cdot \Vert$, because in the 
strict mathematical sense (\ref{enorm}) is not a norm, as it can become negative. Keeping this in mind, we 
return to the Jordan chain (\ref{j1}), (\ref{j2}). According to (\ref{vdf}), the system is solved by the two functions 
$u$ and $v=u_\lambda$. Let us now calculate the Wronskian of these two functions. To this end, we first 
find the Wronskian's derivative by taking into account both equations of the Jordan chain, obtaining
\begin{eqnarray}
\frac{\partial W_{u,u_\lambda}}{\partial x}(x,\lambda) &=& \nonumber \\[1ex]
& & \hspace{-3cm} =~u(x,\lambda)~(u_\lambda)_{xx}(x,\lambda)-
u_{\lambda}(x,\lambda)~u_{xx}(x,\lambda) \nonumber \\[1ex]
& & \hspace{-3cm} =~u(x,\lambda)
 \left\{
 -\left[\lambda-V(x,\lambda) \right] u_\lambda(x,\lambda)
-\left[\lambda-V(x,\lambda) \right]_\lambda u(x,\lambda)
\right\} +u_{\lambda}(x,\lambda) 
\left[\lambda-V(x,\lambda) \right] u(x,\lambda)
\nonumber \\[1ex]
& & \hspace{-3cm} =~-\left[1-V_\lambda(x,\lambda) \right] u^2(x,\lambda). \nonumber
\end{eqnarray}
If we integrate this relation, we obtain the following representation of the Wronskian
\begin{eqnarray}
\int\limits_{x_0}^x \left[1-V_\lambda(t,\lambda) \right] u^2(t,\lambda)~dt &=& W_{u,u_\lambda}(x_0,\lambda)-W_{u,u_\lambda}(x,\lambda), 
\label{eint}
\end{eqnarray}
for an arbitrary $x_0 \in D$. We observe that (\ref{eint}) is a generalization of (\ref{integral}) when $V$ depends on the parameter $\lambda$. In the next step we make the 
following observation: if the function $u$ in 
(\ref{enorm}) is real-valued, then we can immediately match the latter relation with the normalization integral for 
systems that feature energy-dependent potentials. Setting $D=(x_0,x)$, we get
\begin{eqnarray}
N(u) &=& W_{u,u_\lambda}(x_0,\lambda)-W_{u,u_\lambda}(x,\lambda).  \label{normw}
\end{eqnarray}
This identity says that the normalization integral (\ref{enorm}) can be found entirely through calculating derivatives. As mentioned 
before, this is important in applications, because derivatives are generally much easier to find than integrals. 
Furthermore it is important to point out that the identities (\ref{eint}) and (\ref{normw}) neither require 
the function $u$ to be a bound state solution, nor does the associated potential $V$ have to admit such 
bound states. Both of the latter identities are valid independent of any physical meaning, that is, they hold 
for any solution of (\ref{j1}) and for any potential function, as long as the integrals exist. Before we present an example, 
let us comment on the case of a complex-valued function $u$. Since (\ref{eint}) does not contain the absolute value, it is 
not directly applicable. Instead, we can split the 
integral by
\begin{eqnarray}
N(u) &=& \int\limits_D \left[1-V_\lambda(t,\lambda) \right] \mbox{Re}\left[u(t,\lambda) \right]^2 dt+
\int\limits_D \left[1-V_\lambda(t,\lambda) \right] \mbox{Im}\left[u(t,\lambda) \right]^2 dt,\nonumber
\end{eqnarray}
note that Re and Im represent the real- and imaginary part of $u$, respectively. Now, each of these two 
integrals can be identified with our formula (\ref{eint}), provided $u$ is replaced by its real- and imaginary part.

\subsubsection{Energy-dependent harmonic oscillator system} 

Let us now look at a model that features an 
energy-dependent harmonic oscillator potential. Our problem is governed by the following 
boundary-value problem
\begin{eqnarray}
u_{xx}(x,\lambda)+\left(\lambda-\lambda~x^2 \right) u(x,\lambda) &=& 0,~~~(x,\lambda) \in \mathbb{R}^2 
\label{bvp1} \\[1ex]
\lim\limits_{|x| \rightarrow \infty} u(x,\lambda) &=& 0,~~~\lambda \in \mathbb{R}. \label{bvp2}
\end{eqnarray}
This problem is exactly solvable and admits an infinite discrete spectrum $(\lambda_n)$ for an 
associated solution set $(u_n)$, $n$ a nonnegative integer \cite{formanek}, of the form 
\begin{eqnarray}
\lambda_n ~=~ (2~n+1)^2 \quad \qquad u_n(x,\lambda_n) ~=~ \exp\left[-\frac{1}{2}~(2~n+1)~x^2\right] 
H_n \left(\sqrt{2~n+1}~x \right), \label{solbvp}
\end{eqnarray}
where $H_n$ stands for the Hermite polynomial of order $n$ \cite{abram}. For the sake of simplicity, we will 
restrict ourselves to the simplest case $n=0$. We extract from (\ref{solbvp}) 
\begin{eqnarray}
\lambda_0 ~=~ 1 \qquad \qquad \qquad u_0(x,\lambda_0) ~=~ \exp\left(-\frac{1}{2}~x^2 \right). \nonumber
\end{eqnarray}
We will now determine the norm of this function $u_0$ by means of the integral (\ref{enorm}). Afterwards, the 
result is verified by using the differential approach (\ref{normw}). 
The normalization integral (\ref{enorm}) of this function can be evaluated in a straightforward manner 
\begin{eqnarray}
N(u_0) &=& \int\limits_{\mathbb{R}} \left[1-(\lambda~x^2)_\lambda \right]_{\mid \lambda=\lambda_0} 
\exp\left(-x^2 \right) dx ~=~
\int\limits_{\mathbb{R}} (1-x^2)~ \exp\left(-x^2 \right)~dt ~=~ \frac{\sqrt{\pi}}{2}. \label{normc}
\end{eqnarray}
Let us now demonstrate that the same result can be found using our formula (\ref{normw}). First, we must 
determine the partial derivative $(u_0)_\lambda$. This involves a little trick, because we are not given a 
solution of (\ref{bvp1}) in terms of $\lambda$. But since the relation between $\lambda$ (or $\lambda_n$) 
and $n$ in (\ref{solbvp}) is differentiable with differentiable inverse due to nonnegativity of both 
$\lambda_n$ and $n$, we can interpret $n$ as a continuous variable and use the chain rule to obtain
\begin{eqnarray}
(u_0)_\lambda(x,\lambda) &=& \left[\frac{\partial u_n}{\partial n}(x,\lambda_n)~\frac{\partial n}{\partial \lambda_n}
(\lambda_n) \right]_{\mid n=0} \nonumber \\[1ex]
&=& \left[ \frac{\partial u_n}{\partial n}(x,\lambda_n)~\frac{1}{\frac{\partial \lambda_n}{\partial n}(n)} \right]_{\mid n=0} 
\nonumber \\[1ex]
&=& -\frac{1}{4}~x^2~\exp\left(-\frac{1}{2}~x^2 \right) +\frac{1}{4}~\exp\left(-\frac{1}{2}~x^2 \right) 
\left[\frac{\partial H_n}{\partial n}\left(\sqrt{2~n+1}~x \right)\right]_{\mid n=0}  
. \nonumber
\end{eqnarray}
We incorporate this into the Wronskian in (\ref{normw}) and arrive after some simplification at the following result
\begin{eqnarray}
W_{u_0,(u_0)_\lambda} &=&  - \frac{1}{2}~\exp\left(-x^2 \right)
\left[x-H_{-1}(x)\right]. \nonumber
\end{eqnarray}
In the final step we plug this Wronskian into formula (\ref{normw}) for our norm. Since the numbers $x_0$ and 
$x$ correspond to negative and positive infinity, respectively, we must apply limits. This yields
\begin{eqnarray}
N(u_0) &=& 
\lim\limits_{x_0 \rightarrow -\infty} W_{u,u_\lambda}(x_0,\lambda) -\lim\limits_{x \rightarrow \infty} W_{u,u_\lambda}(x,\lambda) \nonumber  \\[1ex]
&=&  - \lim\limits_{x_0 \rightarrow -\infty} \frac{1}{2}~\exp\left(-x_0^2 \right)
\left[x_0-H_{-1}(x_0)\right] + \lim\limits_{x \rightarrow \infty} 
\frac{1}{2}~\exp\left(-x^2 \right)
\left[x-H_{-1}(x)\right]\nonumber  \\[1ex]
&=&  - \lim\limits_{x_0 \rightarrow -\infty} \frac{1}{2}~\exp\left(-x_0^2 \right)
\left[x_0-H_{-1}(x_0)\right] \nonumber  \\[1ex]
&=&  \frac{\sqrt{\pi}}{2}. \nonumber
\end{eqnarray}
This result coincides with (\ref{normc}), as expected. At first sight it seems that the calculations using 
formula (\ref{normw}) are cumbersome when compared to the simple integration in (\ref{normc}). This 
impression is merely due to the simplicity of the present example. If we choose a more general solution of 
(\ref{bvp1}), resolving the integral in (\ref{normc}) can become very difficult. At the same time, the use of  
our formula (\ref{normw}) will also result in complicated calculations, but it involves only differentiation.

\section{Conclusions}

In this work we constructed a relationship between second-order Jordan chains represented in integral and differential forms. This relationship, in a general scenario, allows to find certain integrals involving functions that are solutions of Schr\"odinger equations. We applied this relationship to obtain conditions to generate regular potentials through the confluent SUSY algorithm using the differential representation, to illustrate we used the particle in a box and the radial oscillator systems and in both cases it was shown how to find integrals related to the involved special functions of each system entirely through derivatives. It is worth mention that the exactly solvable SUSY partner potentials of the radial oscillator presented in this work as example had not been reported. Also, since the limits of the found integrals are not fixed, they can be used to calculate probabilities in any arbitrary interval inside the domain of definition of the potential. Moreover, the relationship was used to find normalization constants of wave functions of energy dependent potentials in quantum mechanics, it was exemplified with the ground state of an energy dependent harmonic oscillator. The examples explored in this article are far from being exhaustive but have rather exemplary character since for every quantum system integrals with different special functions can be analyzed

\section*{Acknowledgments}
ACA acknowledges CONACYT fellowship 207577.

\end{sloppypar}
\end{document}